\documentclass[conference]{IEEEtran}
\IEEEoverridecommandlockouts
\def\revisionDiff{0}

\newcommand{\drevised}[1]{\sout{#1}}

\if\revisionDiff0

    \renewcommand{\drevised}[1]{}
\fi

\newcommand\malurl[1]{\href{notalink}{{\nolinkurl{#1}}}}
\newcounter{finding}
\newcommand{\ignore}[1]{}

\usepackage{todonotes}

\usepackage{CJKutf8}
\usepackage{amsmath}
\usepackage{graphicx}
\usepackage[caption=false,font=footnotesize]{subfig}
\usepackage{multirow}
\usepackage{fancyvrb}
\usepackage{threeparttable}

\usepackage[
    colorlinks,
    linkcolor=blue,
    anchorcolor=blue,
    citecolor=blue,
    urlcolor=blue
]{hyperref}

\usepackage{xurl}
\usepackage{array}
\usepackage{algorithmic}
\usepackage{booktabs}
\usepackage{xcolor}
\usepackage{CJKutf8}
\usepackage[normalem]{ulem}
\usepackage[numbers]{natbib}
\usepackage{enumitem}

\usepackage{tcolorbox}
\tcbuselibrary{breakable}
\tcbuselibrary{skins}
\newtcolorbox{promptbox}[1][]{
  colback=white!95!gray,    
  boxrule=0.8pt,            
  rounded corners,          
  title={#1},               
  fonttitle=\bfseries,      
  left=8pt, right=8pt,    
  top=4pt, bottom=4pt,       
  breakable,                
}
\begin{document}

    \title{Detecting and Understanding Illicit Online Promotion with In-Context Learning}
\title{One Prompt to Rule Them All: Cross-Platform Illicit Promotion Detection with In-Context Learning}
\title{Seeing the Unseen: Rethinking Illicit Promotion Detection with In-Context Learning}

\author{
\IEEEauthorblockN{Sangyi Wu\IEEEauthorrefmark{1}, Junpu Guo\IEEEauthorrefmark{1}, Xianghang Mi\IEEEauthorrefmark{1}\IEEEauthorrefmark{2}\thanks{Corresponding to \href{mailto:xianghangmi@gmail.com}{xianghangmi@gmail.com}.}}
\IEEEauthorblockA{\IEEEauthorrefmark{1}University of Science and Technology of China, Hefei, China}
\IEEEauthorblockA{\IEEEauthorrefmark{2}Monash University, Melbourne, Australia}
}
    \maketitle
    
    \begin{abstract}
%
%
Illicit online promotion is a persistent, cross-platform threat that continuously evolves to evade detection. Existing moderation systems, however, remain tethered to platform-specific supervision and static taxonomies—a reactive paradigm that struggles to generalize across domains, adapt to emerging categories, or uncover novel threats before they proliferate.

This paper presents a systematic study of In-Context Learning (ICL) as a unified framework for illicit promotion detection across heterogeneous platforms. Through rigorous analysis of prompt design, we establish that properly configured ICL achieves performance comparable to fine-tuned models using \textbf{22x fewer labeled examples}. More importantly, we demonstrate three capabilities that fundamentally shift the moderation paradigm: (1) \textit{seeing the unseen}—ICL generalizes to entirely new illicit categories without any category-specific demonstrations, incurring a performance drop of less than 6\% for over half of the 12 evaluated categories; (2) \textit{autonomous discovery}—a novel two-stage pipeline distills over 2,900 free-form labels into coherent taxonomies, surfacing \textbf{eight previously undocumented illicit categories} including usury and illegal immigration; and (3) \textit{cross-platform generalization}—deployed on 200,000 real-world samples from search engines and Twitter without any platform adaptation, ICL achieves 92.6\% accuracy, with 61.8\% of its uniquely flagged samples corresponding to borderline or obfuscated promotional content missed by existing detectors.

Our findings position ICL as a new paradigm for content moderation—one that combines the precision of specialized classifiers with uniquely powerful capabilities: cross-platform generalization without adaptation, autonomous discovery of emerging threats, and dramatic reductions in labeled data requirements. By shifting from static, taxonomy-driven supervision to inference-time reasoning, ICL offers a path toward moderation systems that are not just reactive but proactively adaptive.
The source code of this work is available at \url{https://github.com/ChaseSecurity/illicit-icl}.
    \end{abstract}
    
    \begin{CJK*}{UTF8}{gbsn}
        

\section{Introduction}
\label{sec:intro}
The internet's underground economy thrives on a persistent, shape-shifting threat: illicit online promotion. From drug trafficking and counterfeit goods to illegal gambling and financial fraud, this content permeates search engines, social media, and messaging apps~\cite{leontiadisMeasuringAnalyzingSearchRedirection2011b,wangDemystifyingLocalBusiness2022a,zhaUnderstandingCrossPlatformReferral2024a,liaoSeekingNonsenseLooking2016d,yangCasinoRoyaleDeep2019c}. While its societal harm is undeniable, what makes this threat particularly insidious is its \textbf{cross-platform, adaptive nature}. Adversaries do not confine themselves to a single service; they strategically distribute their operations—using search engine optimization (SEO) to attract victims, social media for amplification, and encrypted messaging for final transactions~\cite{wuIdentifyingLinkFarm2005c,liaoCharacterizingLongtailSEO2016,wang2025twitter,zhaUnderstandingCrossPlatformReferral2024a}. This constant migration exploits the structural and linguistic gaps between platforms, creating a moving target that renders traditional, platform-specific defenses brittle~\cite{wangCloakDaggerDynamics2011b,invernizziCloakVisibilityDetecting2016b,liaoSeekingNonsenseLooking2016d}.

For years, machine learning-based moderation has fought this battle with a reactive playbook: train a classifier on labeled data for a specific platform, then hope it holds up. But in the open world of online threats, this approach is failing. The emergence of new illicit categories, intentional linguistic obfuscation, and distribution shifts across platforms all cause the performance of fine-tuned models to degrade precipitously. The result is a costly, unsustainable cycle of continuous data annotation, retraining, and manual intervention—a paradigm ill-suited for a problem defined by constant evolution. What we need is not just a better classifier, but a fundamental shift: from a \textit{closed-world} assumption of fixed categories to an \textit{open-world} capability for generalization and discovery.

{In-Context Learning (ICL)} with Large Language Models (LLMs) offers precisely this paradigm shift~\cite{brownLanguageModelsAre2020}. By conditioning on a handful of examples provided in a prompt, ICL enables models to learn new tasks at inference time without any parameter updates. This unlocks the potential for a unified, platform-agnostic moderation framework that can leverage the vast semantic knowledge learned during pretraining to adapt on the fly.
However, deploying ICL in the adversarial, multilingual, and heterogeneous landscape of illicit promotion is not straightforward. It forces us to confront a series of open questions. How do we design prompts that are robust to adversarial text? Can ICL truly generalize to entirely new, unseen categories of illicit activity without retraining? And can we move beyond simple classification to use ICL as a tool for autonomously \textit{discovering} emerging threats from raw, unlabeled data?

This paper provides the first systematic study to answer these questions, transforming ICL from a promising technique into a practical framework for next-generation content moderation. Our investigation spans controlled experiments to large-scale, real-world deployments, with four primary contributions.

\vspace{4pt}
\noindent\textbf{A blueprint for effective ICL in security.} Through rigorous analysis of prompt configurations—demonstration quantity, selection strategies, and label design—we uncover the inductive biases and critical configurations that make ICL robust for adversarial content. We show that with optimal design, ICL achieves performance comparable to fine-tuned models using \textbf{22x fewer labeled examples}, and that explicit semantic labels are essential—without them, the false positive rate jumps from 8\% to 42\%.

\vspace{4pt}
\noindent\textbf{The ability to see the unseen.} In a stringent open-world evaluation, we demonstrate that ICL generalizes effectively to entirely new illicit categories without any category-specific examples, relying instead on shared semantic characteristics of illicit intent. For more than half of the 12 evaluated categories, excluding the target category from demonstrations incurs a performance drop of less than 6\% relative to the category-included setting.

\vspace{4pt}
\noindent\textbf{Autonomous discovery of emerging threats.} Moving beyond static taxonomies, we design a novel two-stage pipeline where ICL first generates open-ended category labels for unlabeled data, then self-consolidates them into coherent, novel threat taxonomies. This process distills over 2,900 free-form labels into meaningful clusters and autonomously surfaces \textbf{eight distinct new illicit categories}—including usury, illegal immigration, and software piracy—absent from prior studies.

\vspace{4pt}
\noindent\textbf{Validation in the wild.} On large-scale, cross-platform data (200,000 samples from search engines and Twitter), our ICL framework outperforms platform-specific baselines without any adaptation, achieving 92.6\% accuracy. Crucially, 61.8\% of samples uniquely flagged by ICL correspond to valid borderline or heavily obfuscated promotional content missed by existing detectors, positioning ICL as a powerful high-recall discovery funnel for hybrid moderation systems.

\vspace{8pt}
Overall, this work reframes illicit promotion detection as an adaptive, open-world challenge and demonstrates that in-context learning provides a viable framework for addressing it. By shifting from static, taxonomy-driven supervision toward inference-time reasoning, we pave the way for moderation systems that can keep pace with the very threats they aim to stop. To support reproducibility and facilitate future research, we release our code, prompt templates, and curated benchmark datasets at \url{https://github.com/ChaseSecurity/illicit-icl} under permissive licensing.

\section{Related Works}
\label{sec:related}
\subsection{Illicit Online Promotion}

Illicit online promotion represents a persistent and evolving threat where miscreants leverage the internet's vast reach to promote illicit goods and services. To maximize visibility and traffic, adversaries have developed a diverse arsenal of techniques targeting search engines, legitimate web infrastructure, and online social networks.

Early research into illicit promotion largely focused on blackhat search engine optimization (SEO). Wu and Davison~\cite{wuIdentifyingLinkFarm2005c} identified the use of link farms, where densely connected networks of spam pages are created to manipulate graph-based ranking algorithms. As search engines improved their defenses, attackers evolved to adopt cloaking techniques, which involve serving benign content to search engine crawlers while displaying malicious content to human visitors. Wang et al.~\cite{wangCloakDaggerDynamics2011b} characterized this dynamic, and Invernizzi et al.~\cite{invernizziCloakVisibilityDetecting2016b} further formalized it, developing systems to detect when machines browse a different web than humans.

A parallel strategy involves compromising legitimate, high-reputation websites to exploit their domain authority. Leontiadis et al.~\cite{leontiadisMeasuringAnalyzingSearchRedirection2011b} analyzed search-redirection attacks, demonstrating how compromised high-value sites (e.g., .edu domains) were injected with code to divert search traffic to illicit online pharmacies. Building on this, Liao et al.~\cite{liaoSeekingNonsenseLooking2016d} proposed detection methods based on identifying the semantic inconsistency between such injected promotional content and the victim site's original context.

Beyond direct compromise, adversaries abuse public infrastructure to host and promote content. Liao et al.~\cite{liaoCharacterizingLongtailSEO2016} uncovered long-tail SEO spam hosted on reputable cloud platforms, which used doorway pages to monetize traffic through affiliate programs. Similarly, the abuse of local business listings has become a critical vector. Huang et al.~\cite{huangPinningAbuseGoogle2017} and Wang et al.~\cite{wangDemystifyingLocalBusiness2022a} investigated how miscreants poison local search services (e.g., Google Maps) with fake listings to promote affiliate fraud and illicit drugs. More recently, Lin et al.~\cite{linMAWSEOAdversarialWiki2024a} demonstrated adversarial Wiki search poisoning, where attackers manipulate open-collaboration platforms to boost the ranking of illicit content. Distinct from these injection-based methods, Wu et al.~\cite{wu2024reflectedsearchpoisoningillicit} identified reflected search poisoning, a technique where attackers abuse legitimate URL reflection schemes on benign websites to index illicit promotion texts without compromising the server.

With the rise of social media, illicit promotion has expanded beyond search engines. Wang et al.~\cite{wang2025twitter} performed a large-scale analysis of illicit promotion on Twitter, identifying millions of posts across diverse categories such as data leakage, pornography, and contraband. Yang et al.~\cite{yangCasinoRoyaleDeep2019c} examined the ecosystem of illegal online gambling, revealing a complex profit chain involving promotion via social apps, blackhat SEO, and third-party payment abuse. Furthermore, modern illicit promotion often exhibits complex cross-platform behaviors. Zha et al.~\cite{zhaUnderstandingCrossPlatformReferral2024a} revealed a cross-platform referral traffic ecosystem, where drug dealers utilize video platforms like TikTok and YouTube to attract potential buyers before redirecting them to encrypted messaging apps or storefronts on other platforms to consummate the sale.

Collectively, these studies illustrate an arms race where illicit promotion strategies continuously shift across platforms and modalities. The inherent heterogeneity of these distribution channels poses a significant and ongoing challenge for traditional, static detection systems.

\subsection{In-Context Learning}

Large Language Models (LLMs) have demonstrated a remarkable emergent ability known as In-Context Learning (ICL), which allows models to perform novel tasks by conditioning on a few input-output pairs provided in the prompt, without updating any model parameters~\cite{brownLanguageModelsAre2020}. Unlike traditional supervised fine-tuning, ICL enables rapid adaptation to downstream tasks through inference alone, fundamentally shifting the paradigm of NLP system deployment. 
ICL is distinct from related concepts like prompt learning~\cite{liuPretrainPromptPredict2023} and standard few-shot learning~\cite{wangGeneralizingFewExamples2020}. Its defining characteristics offer several key advantages: it supports interpretable human-LLM interaction through natural language demonstrations~\cite{brownLanguageModelsAre2020}; its analogical reasoning mechanism aligns with human cognitive processes~\cite{winstonLearningReasoningAnalogy1980}; and its training-free nature significantly reduces the adaptation cost for deploying large models in real-world scenarios~\cite{sunBlackBoxTuningLanguageModelasaService2022}.

To maximize the effectiveness of ICL, the selection and presentation of demonstration examples are critical. Empirical research has established several best practices. While increasing the number of demonstrations generally improves performance~\cite{agarwalManyShotInContextLearning2024}, the marginal gains often diminish beyond a certain point~\cite{brownLanguageModelsAre2020, minRethinkingRoleDemonstrations2022b}. More importantly, regarding demonstration selection, compared to the traditional approach of using a fixed set of few-shot demonstrations for all queries, dynamically retrieving customized demonstrations tailored to each specific query has yielded significant performance gains \cite{luoDrICLDemonstrationRetrievedIncontext2023, yeCompositionalExemplarsIncontext2023a}. To enable this, unsupervised selection methods have been developed, utilizing metrics such as k-nearest neighbor retrieval based on sentence embeddings~\cite{liuWhatMakesGood2022, qinInContextLearningIterative2024a}, mutual information~\cite{sorensenInformationtheoreticApproachPrompt2022}, perplexity~\cite{gonenDemystifyingPromptsLanguage2023}, and various information-theoretic scores~\cite{wuSelfAdaptiveInContextLearning2023a, liFindingSupportExamples2023a}. Beyond the selection and quantity of examples, the order in which they appear also significantly impacts model behavior, with performance being highly sensitive to prompt ordering~\cite{luFantasticallyOrderedPrompts2022b} and a recency bias often observed in model predictions~\cite{zhaoCalibrateUseImproving2021c}.

The flexibility of ICL has driven its adoption in content moderation and security applications, where data scarcity and evolving threats pose challenges for supervised models. Wang and Chang~\cite{wangToxicityDetectionGenerative2022} utilized generative prompt-based inference for zero-shot toxicity detection, demonstrating advantages in handling long-tail phenomena. Recent frameworks have integrated ICL into more comprehensive moderation pipelines. For example, LLM-Mod~\cite{kollaLLMModCanLarge2024} leverages LLMs to identify rule violations in online communities, employing multi-step prompting to enhance reasoning about guidelines. Similarly, Zhang et al.~\cite{zhangEfficientToxicContent2024} proposed bootstrapping LLMs via a Decision-Tree-of-Thought prompting strategy to detect toxic content and distill rationales into smaller models. In the realm of factuality and fairness, Zhang et al.~\cite{zhangInterpretableUnifiedLanguage2023} introduced a unified checking framework that combines fact generation with ethical classification prompts. 

While these works demonstrate the efficacy of ICL for specific moderation tasks, our work extends this paradigm to the detection of illicit online promotion, a distinct challenge characterized by cross-platform distribution and adversarial adaptation.

\section{Experimental Setup}
\label{sec:setup}
\subsection{Datasets}
To comprehensively evaluate the efficacy of in-context learning (ICL) for detecting and understanding illicit online promotion, we construct two purpose-built datasets: a Binary Dataset for illicit vs. benign classification and a fine-grained Multiclass Dataset for identifying specific illicit categories. These datasets are derived from two large-scale, real-world sources to ensure diversity and practical relevance.

\noindent\textbf{Data Sources.}
The data employed in this study draw upon two existing research initiatives examining illicit promotion across distinct online platforms, thereby capturing varied distribution channels and linguistic contexts.
The first source consists of Illicit Promotion Texts (IPTs) collected via Reflected Search Poisoning (RSP) techniques, as detailed in the work by Wu et al. \cite{wu2024reflectedsearchpoisoningillicit}. This dataset comprises over 11 million entries spanning 14 illicit categories—including drug trading, data theft, and hacking services—generated by manipulating high-ranking websites to poison search engine results. It thus represents a covert, web-based promotional vector.
The second source derives from a study of illicit promotion within Online Social Networks (OSNs), specifically Twitter, as presented by Wang et al. \cite{wang2025twitter}. This collection contains 12 million Posts of Illicit Promotion (PIPs) across 10 illicit categories such as drugs, gambling, and weapon sales, posted in five major natural languages. It reflects overt promotional activities within public social media discourse.

\noindent\textbf{Dataset Integration and Category Alignment.}
The original sources used different categorization schemas. To create a coherent multiclass dataset, we performed a semantic correlation and unification of categories across the two works. This process consolidated overlapping categories (e.g., "Illegal Drug Sales" from IPTs and "drug" from PIPs) and resolved naming discrepancies, resulting in a unified taxonomy of 12 distinct illicit categories. The mapping is detailed in Table~\ref{tab:category}.

\begin{table}[htbp]
    \centering
    \caption{\textbf{Taxonomy unification mapping.} Alignment of heterogeneous categories from Twitter and Search Engine datasets into the unified taxonomy used in this study.}
    \label{tab:category}
    \begin{tabular}{lll}
        \toprule
        Unified Category & Source: Twitter\cite{wang2025twitter} & Source: Search Engine\cite{wu2024reflectedsearchpoisoningillicit} \\
        \midrule
        porn & porn & Illegal Sex \\
        surrogacy & surrogacy & Illegal Surrogacy \\
        gambling & gambling & Gambling \\
        drug & drug & Illegal Drug Sales \\
        weapon & weapon & Illegal Weapon Sales \\
        data-theft & data\_leakage & Data Theft \\
        money-laundry & money-laundry & Money Laundering \\
        advertisement & crowdturfing & Black Hat SEO \& Adv. \\
        counterfeit & fake\_document & Fake Account \\
                    &               & Fake Certificate \\
                    &               & Counterfeit Goods \\
        hacking & -- & Hacking Service \\
        fraud & -- & Financial Fraud \\
        others & harassment & Others \\
        \bottomrule
    \end{tabular}
\end{table}

\noindent\textbf{Balanced Dataset Construction.}
To mitigate potential model bias, a balanced dataset was constructed for the binary classification task distinguishing illicit from benign content. The original data comprised 9,352 illicit and 4,395 benign instances collected from two distinct sources. Stratified resampling was applied to achieve both class balance and source diversity. 
The resulting Binary Dataset consists of 5,600 samples, maintaining an equal distribution across labels and sources. Specifically, the dataset contains 2,800 illicit and 2,800 benign samples, with each label containing 1,400 samples from Twitter and 1,400 from search engine data. Then, it was divided into a training set and a test set in a 4:1 ratio.

For the multiclass categorization task, a separate balanced dataset was assembled across 12 unified illicit categories and one benign class. Each of these 13 categories was resampled to contain 500 instances, yielding a total of 6,500 samples. To preserve source diversity where feasible, a 1:1 source ratio was targeted; however, for categories underrepresented in a particular source (such as “hacking” which was absent from Twitter), samples were drawn exclusively from the available source. The final Multiclass Dataset contains 3,264 samples from Twitter and 3,236 from search engines. Linguistically, the dataset is diverse, predominantly composed of Chinese (68.3\%) and English (19.9\%), with additional representation from Japanese, Korean, Thai, and other languages.

\subsection{Models}

To ensure a comprehensive and representative evaluation of In-Context Learning (ICL) capabilities across different model architectures and scales, we select a diverse suite of modern, open-source decoder-only large language models (LLMs). The models are: 
\textbf{Llama}\cite{grattafiori2024llama3herdmodels} (llama-3.1-8B\footnote{https://huggingface.co/meta-llama/Llama-3.1-8B}), 
\textbf{Mistral}\cite{jiang2023mistral7b} (mistral-7b-instruct-v0.2\footnote{https://huggingface.co/mistralai/Mistral-7B-Instruct-v0.2}), 
\textbf{Phi3-Small}\cite{abdin2024phi3technicalreporthighly} (phi-3-small-128k-instruct\footnote{https://huggingface.co/microsoft/Phi-3-small-128k-instruct}), 
\textbf{Phi3-Mini}\cite{abdin2024phi3technicalreporthighly} (phi-3-mini-128k-instruct\footnote{https://huggingface.co/microsoft/Phi-3-mini-128k-instruct}), 
\textbf{Qwen}\cite{qwen2025qwen25technicalreport} (qwen2.5-7b-instruct\footnote{https://huggingface.co/Qwen/Qwen2.5-7B-Instruct}), 
and \textbf{Gemma}\cite{gemmateam2024gemma2improvingopen} (gemma-2b-it\footnote{https://huggingface.co/google/gemma-2b-it}).





\section{Analysis of Prompt Configurations}
\label{sec:effective}
Before deploying ICL for large-scale detection, it is essential to understand how different prompt engineering factors interact with the specific nuances of illicit promotion. In this section, we systematically deconstruct the ICL configuration to identify the optimal setup. We examine four critical dimensions: the quantity of demonstrations, selection strategies, the underlying model architecture (instruction tuning), and the semantic influence of label verbalization. These analyses serve as the foundation for the performance evaluations in subsequent sections.

\subsection{Impact of Demonstration Quantity}
\label{sec:effective_shots}
A fundamental question in deploying in-context learning (ICL) is determining the optimal number of demonstration examples. Providing too few may fail to adequately convey the task, while too many may introduce noise, increase computational cost, and potentially exceed the model's effective context window. To investigate this, we evaluated all models with shot counts $k\in[0,2,4,8,16,32,64,128]$ across binary and multiclass tasks. The demonstrations were selected randomly from the training dataset of each task, and we experiment with 10 seeds to calculate averaged scores.

During our experiments, we observed that Gemma could not reliably perform classification, failing to output the predefined labels with a failure rate as high as 58.94\%.. Consequently, we excluded it from further experiments and analysis. For similar reasons, Phi3-mini was also omitted from the multiclass classification task.

\begin{figure}[htbp]
    \centering
    \subfloat[Binary classification.\label{fig:binary_increasing_shots}]{%
        \includegraphics[width=0.36\textwidth]{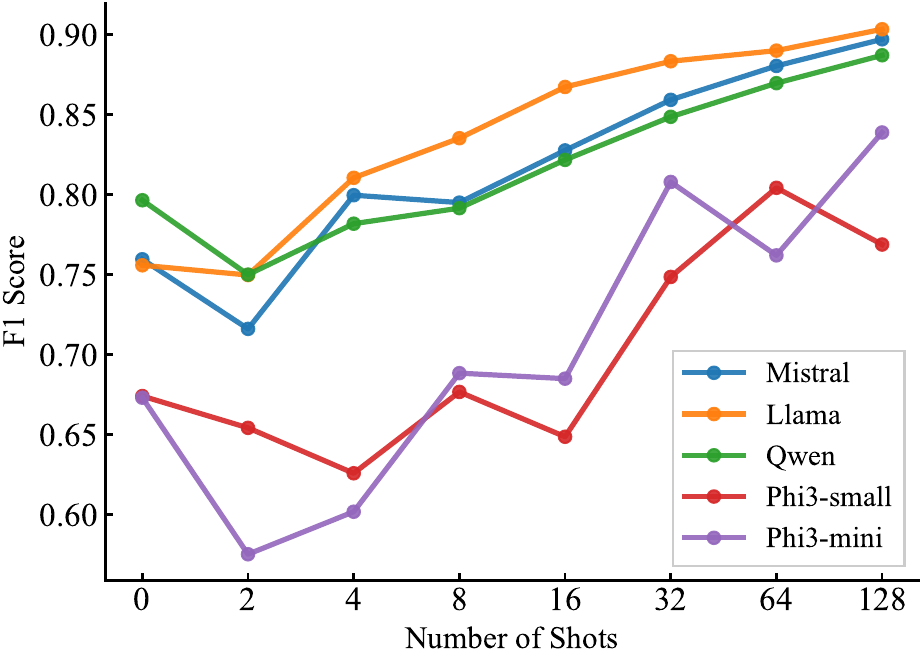}%
    }
    \hfill
    \subfloat[Multiclass classification.\protect\footnotemark\label{fig:multiclass_increasing_shots}]{
        \includegraphics[width=0.36\textwidth]{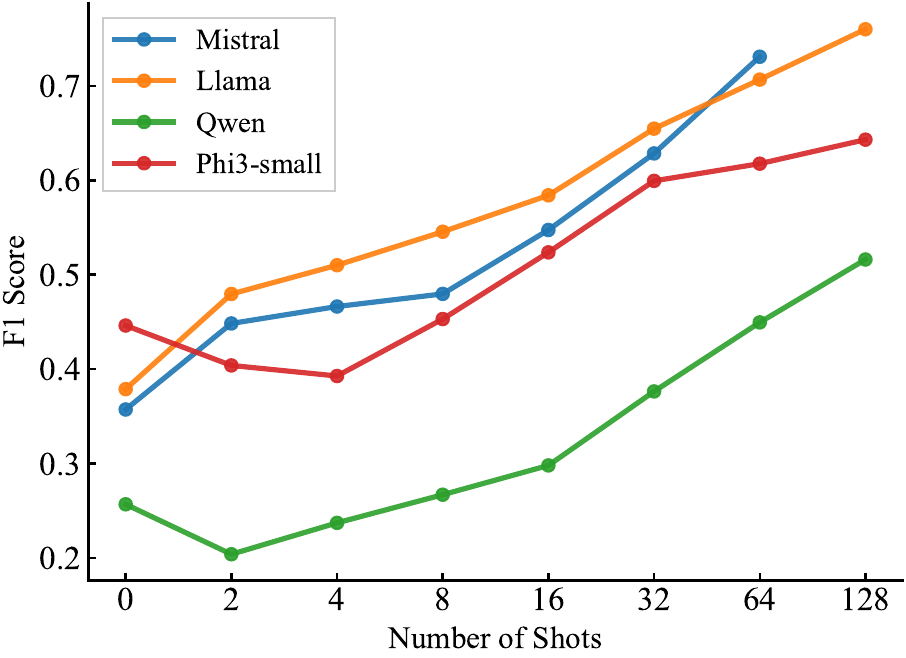}
    }
    \caption{\textbf{Impact of Demonstration Quantity.} F1-scores of different model families as a function of the number of shots for (a) binary and (b) multiclass classification tasks.}
    \label{fig:increasing_shots_comparison}
\end{figure}

\noindent\textbf{Binary Classification.}
As illustrated in Fig.~\ref{fig:binary_increasing_shots}, most models benefit from increasing the number of shots in binary classification, though the extent and saturation point of improvement vary. Models such as Mistral, Llama, and Qwen show steady gains as $k$ rises from 0 to 32, after which performance plateaus.
In contrast, the Phi3 family exhibits less stable behavior. Phi3-small and Phi3-mini display noticeable fluctuations under few-shot conditions, indicating greater sensitivity to the choice of demonstrations. Although they improve with more examples, their progress is less consistent and saturates earlier than other models. Notably, a sharp performance jump occurs between 16 and 32 shots, suggesting that a minimum amount of contextual supervision is needed to activate their classification capability.
Overall, binary classification tasks can be effectively learned with a moderate number of demonstrations (e.g., 16–32 shots), beyond which additional examples yield diminishing returns.

\noindent\textbf{Multiclass Classification.}
Fig.~\ref{fig:multiclass_increasing_shots} reveals a distinctly different trend for multiclass classification. Unlike in the binary setting, all models exhibit a stronger and more consistent reliance on the number of shots, with no clear plateau even at 128 shots. This suggests that multiclass tasks place greater demands on in-context learning, requiring more examples to adequately capture inter-class decision boundaries.
Among the evaluated models, Llama and Mistral again outperform others across all shot counts, demonstrating nearly linear improvement as $k$ increases. Phi3-small shows moderate gains but consistently lags behind, reflecting its limited ability to generalize across multiple classes from context alone. Qwen performs notably worse in low-shot settings and improves more gradually, although its gap narrows with more shots.
These findings highlight that multiclass classification is substantially more sensitive to the number of in-context examples than binary classification. While adding shots consistently improves accuracy, the lack of a clear saturation point implies that practical deployment must balance accuracy gains against increased inference costs and context-length constraints.


\subsection{Demonstration Selection Strategies}
While increasing the number of demonstrations generally improves model performance, relying solely on random selection overlooks a critical factor: the relevance and quality of each demonstration to the target query. Intuitively, a few highly pertinent examples should offer more actionable context than numerous irrelevant ones. To investigate this systematically, we compare three demonstration selection strategies:

\begin{itemize}
    \item Random: Baseline selection without considering query content.
    \item Lexical-based: Employs the BM25 algorithm\cite{bm25}, a bag-of-words retrieval function that ranks demonstrations based on lexical overlap with the query.
    \item Semantic-based: Utilizes a pre-trained Sentence-Transformer\cite{reimers-2019-sentence-bert} model\footnote{https://huggingface.co/sentence-transformers/paraphrase-multilingual-MiniLM-L12-v2} to encode both queries and demonstrations into dense vector embeddings. Demonstrations are ranked by their cosine similarity to the query in this semantic space.
\end{itemize}

\footnotetext{Due to the context length limit of Mistral, the 128-shot experiment could not be conducted.}

\begin{figure}[htbp]
    \centering
    \subfloat[Binary classification. (Mistral)\label{fig:binary_retrieval_mistral}]{%
        \includegraphics[width=0.24\textwidth]{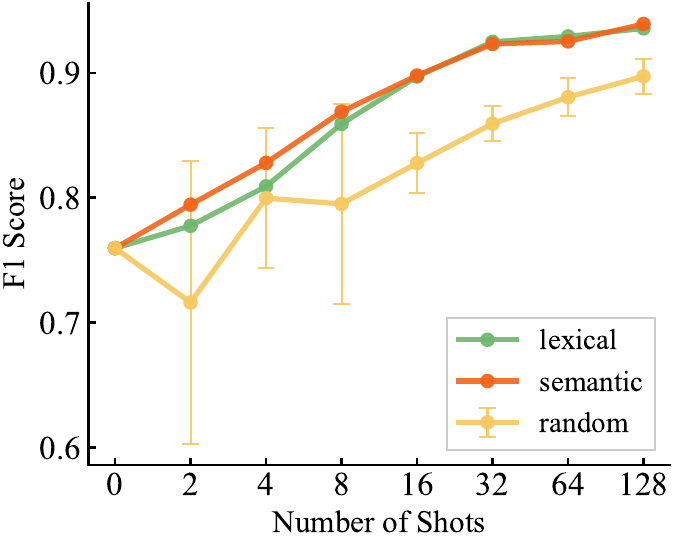}%
    }
    \hfill
    \subfloat[Multiclass classification. (Mistral)\label{fig:multiclass_retrieval}]{%
        \includegraphics[width=0.24\textwidth]{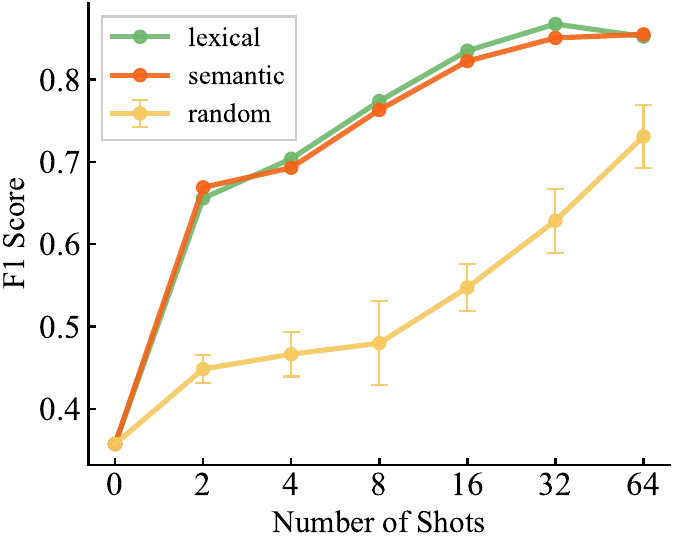}%
    }
    \caption{\textbf{Efficacy of Demonstration Selection Strategies.} Comparison of Random, Lexical (BM25), and Semantic (Embedding-based) selection strategies across varying shot counts using the Mistral model.\protect\footnotemark}
    \label{fig:selection_comparison}
\end{figure}

\noindent\textbf{Binary Classification.}
Fig.~\ref{fig:binary_retrieval_mistral} illustrates the impact of different demonstration selection strategies on binary classification performance. Overall, both lexical-based and semantic-based retrieval consistently outperform random selection across almost all shot settings. This gap is particularly pronounced in the low-shot regime ($k \leq 8$), where random selection exhibits high variance and unstable performance, as reflected by the larger error bars. In contrast, retrieval-based strategies provide more reliable and informative contextual signals, leading to faster performance gains with fewer demonstrations.
Between the two retrieval approaches, semantic-based selection generally achieves the best or comparable performance to lexical-based selection, especially as the number of shots increases. This suggests that capturing semantic similarity beyond surface-level lexical overlap is beneficial for selecting more informative demonstrations. As $k$ grows larger (e.g., $k \geq 32$), the performance gap between strategies gradually narrows, indicating that the negative impact of suboptimal demonstration selection can be partially mitigated by increasing the number of examples.

\noindent\textbf{Multiclass Classification.}
As shown in Fig.~\ref{fig:multiclass_retrieval}, the advantages of informed demonstration selection become even more pronounced in multiclass classification. Both lexical-based and semantic-based strategies significantly outperform random selection across all shot counts, with a substantial margin that persists even at higher values of $k$. Random selection struggles to provide sufficient class-discriminative information, resulting in consistently lower F1 scores and slower improvement as more shots are added.
Semantic-based retrieval again demonstrates slightly stronger or comparable performance relative to lexical-based retrieval, particularly in the mid-to-high shot range. This highlights the importance of semantic alignment between the query and selected demonstrations when dealing with more complex decision boundaries involving multiple classes. 

Overall, these results indicate that demonstration selection strategy plays a critical role in in-context learning, and that semantically informed retrieval is an effective and robust choice for improving classification performance.

\footnotetext{Mistral was selected because it performs best; see the appendix~\ref{app:mistral} for details.}


\subsection{Instruction Tuning and Model Alignment}
As established, the effectiveness of in-context learning is significantly influenced by both the number and the selection of demonstrations. However, these configuration choices act upon a more fundamental element: the underlying large language model (LLM) itself. A critical architectural distinction among LLMs is whether they have undergone instruction tuning, a supervised fine-tuning process designed to align the model's outputs with human instructions and desired response formats. 
To isolate the impact of this property, we conduct a controlled comparison between the instruction-tuned and non-instruction-tuned (base) variants of three model families: Llama, Mistral and Qwen. All experiments employ 32 demonstrations selected via semantic retrieval.

\begin{figure}[htbp]
    \centering
    \includegraphics[width=0.36\textwidth]{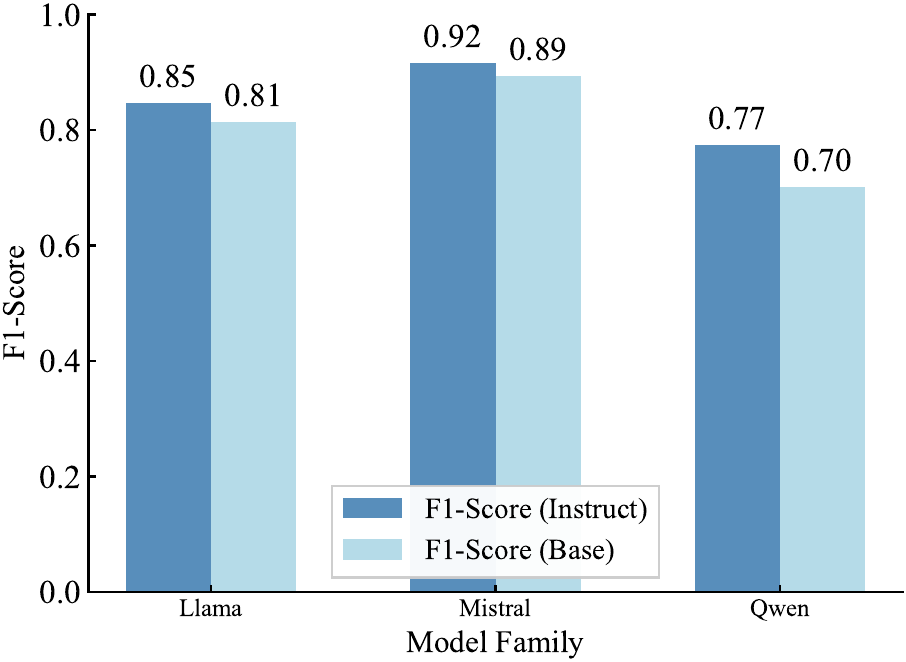}
    \caption{\textbf{Instruction Tuning Impact.} Performance comparison between Instruction-Tuned and Base variants across Llama, Mistral, and Qwen families (32-shot, semantic retrieval).}
    \label{fig:instruct_comparison}
\end{figure}

As illustrated in Fig.~\ref{fig:instruct_comparison}, the instruction-tuned variants consistently outperform their base counterparts across all three families. For both Llama and Mistral, instruction tuning yields clear and stable gains in F1 score, indicating that alignment with human instructions substantially enhances the model's capacity to follow classification prompts and generate well-structured label outputs in the ICL setting. While Qwen exhibits lower overall performance compared to the other models, instruction tuning still provides a noticeable improvement. This suggests that the benefits of instruction tuning are generalizable across model families, even when overall classification capability is constrained by other architectural or training factors.

These results underscore that instruction tuning is a critical enabler of effective in-context learning. Beyond improving raw accuracy, instruction-tuned models show better adherence to task specifications and output constraints, thereby reducing ambiguity in label generation. Consequently, while demonstration quantity and selection strategy are important operational factors, the choice of an instruction-tuned backbone model plays a foundational role in achieving reliable and robust ICL performance.

\subsection{Label Necessity and Semantic Verbalization}
With the optimal demonstration strategy and backbone model established, we now turn to the final component of the prompt: the label verbalization. While instruction-tuned models are better at following rules, the semantic clarity of those rules (i.e., the labels) remains a critical variable. This section investigates two fundamental aspects of label usage: whether explicit labels are indispensable, and how the semantic content of the label words themselves influences task learning.

\begin{figure}[htbp]
\centering
\includegraphics[width=0.36\textwidth]{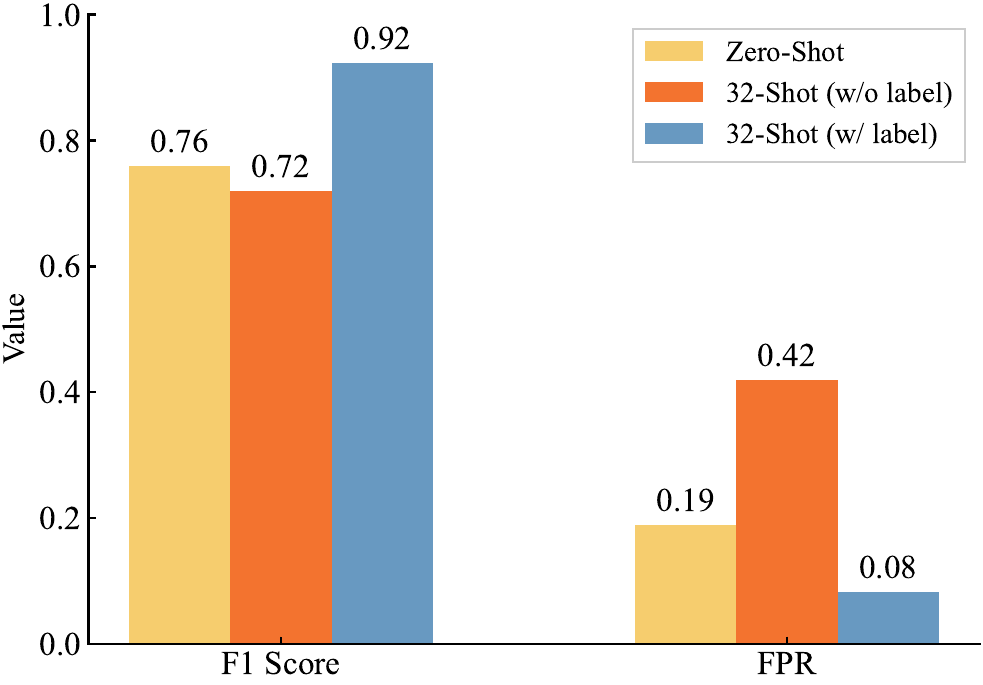}
\caption{\textbf{Necessity of Explicit Labels.} Impact of removing labels from demonstrations compared to Zero-shot and Standard (Labeled) Few-shot baselines on FPR and F1 Score.}
\label{fig:without_label_comparison}
\end{figure}

\noindent\textbf{The Necessity of Explicit Labels.}
To determine whether LLMs can infer task rules from input examples alone, we compare three experimental settings for the Mistral model: zero-shot, 32-shot without labels, and 32-shot with labels. As summarized in Figure~\ref{fig:without_label_comparison}, explicit labels prove critical.

Removing labels from the demonstrations led to a significant performance drop relative to the labeled few-shot setting, and performance was even worse than the zero-shot baseline. Notably, the False Positive Rate (FPR) increased sharply in the no‑label condition. This suggests that without explicit labels, the model cannot reliably infer the intended classification rule from the demonstration texts alone. Instead, it appears to default to a biased strategy, frequently predicting the positive (illicit) class. These results confirm that labels are not merely helpful but essential in the ICL framework for content moderation. They provide the definitive supervisory signal that anchors the model’s interpretation of the demonstration context, preventing degenerate learning outcomes.

\noindent\textbf{Sensitivity to Label Verbalization.}
Given that labels are necessary, we further examine whether their semantic meaning matters. We experiment with three verbalization schemes for the Mistral model: \textit{Original} (benign/illicit), \textit{Inverted} (illicit/benign), and \textit{Abstract} (0/1 or A/B). The results reveal a pronounced difference between binary and multiclass tasks.

\begin{figure}[htbp]
    \centering
    \subfloat[Binary classification.\label{fig:binary_label_name_mistral}]{%
        \includegraphics[width=0.24\textwidth]{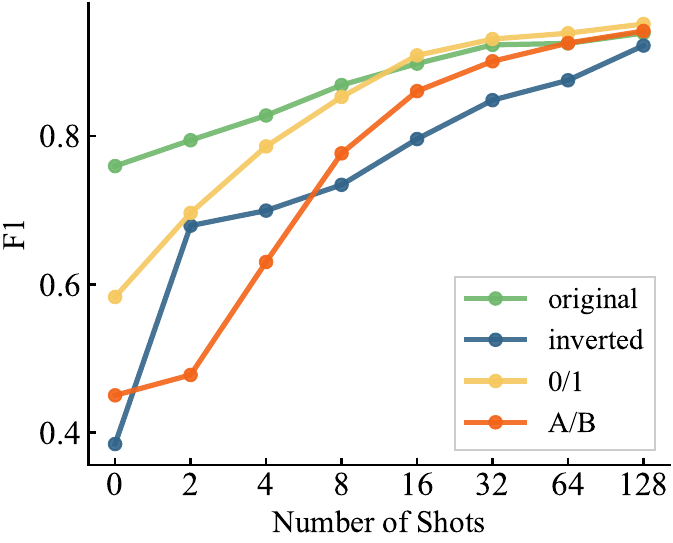}%
    }
    \hfill
    \subfloat[Multiclass classification.\label{fig:multiclass_label_name_mistral}]{%
        \includegraphics[width=0.24\textwidth]{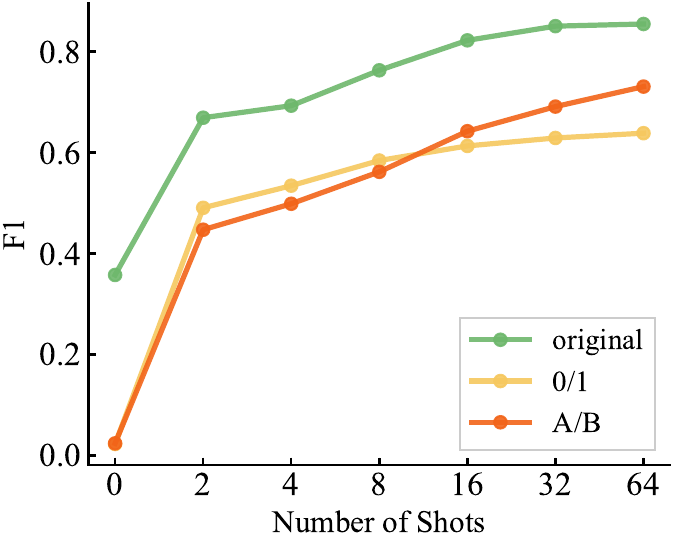}%
    }
    \caption{\textbf{Sensitivity to Label Verbalization.} Comparison of Original (Semantic), Inverted, and Abstract (Symbolic) label mappings for (a) binary and (b) multiclass tasks.}
    \label{fig:label_name}
\end{figure}

For binary classification (Fig.~\ref{fig:binary_label_name_mistral}), in low‑shot regimes ($k<16$), the model relies heavily on its pre‑existing semantic knowledge. The original labels, whose meanings align with the task, perform best. While semantically contradictory (inverted) or meaningless (abstract) labels perform poorly. As the number of shots increases ($k\ge32$), this pattern shifts. The model begins to learn primarily from the input–label mapping within the demonstrations rather than from the intrinsic meaning of the label words. Consequently, the performance of inverted and abstract labels converges with or even slightly surpasses that of original labels. The simplicity of abstract symbols may reduce cognitive load, making the mapping rule easier to learn from abundant examples.

The pattern for multiclass classification (Fig.~\ref{fig:multiclass_label_name_mistral}) is markedly different. Here, semantically meaningful original labels consistently and substantially outperform abstract labels across all shot counts. Although abstract labels still improve over zero‑shot performance, they fail to match original labels. Multiclass classification requires disambiguating multiple fine‑grained categories. Semantic labels (e.g., ‘porn’, ‘gambling’) provide direct, pre‑learned conceptual anchors that help the model quickly associate demonstration content with the correct class. Abstract labels lack this semantic scaffolding, forcing the model to learn a completely arbitrary and more complex mapping from scratch, which proves inefficient within limited context.

\section{Comparative Efficiency and Robustness Analysis}
\label{sec:analyze}
Having identified the optimal configuration for ICL, a fundamental question remains: is this inference-only paradigm truly competitive with traditional training-based methods? In this section, we move beyond configuration to a rigorous benchmarking analysis. We specifically evaluate ICL’s data efficiency against parameter-efficient fine-tuning (LoRA) and probe its mechanistic robustness against positional biases and noise, establishing the theoretical and practical grounds for its deployment.

\subsection{Data Efficiency: ICL vs. Parameter-Efficient Fine-Tuning}
A key advantage of In-Context Learning (ICL) lies in its ability to adapt to downstream tasks without updating model parameters, relying solely on task demonstrations provided within the input prompt. This property makes ICL particularly attractive in scenarios where computational resources or labeled data are limited. In contrast, parameter-efficient fine-tuning (PEFT) methods, such as Low-Rank Adaptation (LoRA), explicitly modify a subset of model parameters using supervised training data. This section seeks to answer a practical and fundamental question: how does the performance of ICL, scaled by the number of in-context demonstrations ($k$-shot), compare with that of LoRA fine-tuning, scaled by the number of training samples ($K$)?

To ensure a fair comparison, we conduct all experiments using Mistral as the base model for both approaches. For ICL, we adopt the best-performing configuration, including instruction-tuned checkpoints and semantic retrieval for demonstration selection. 
For LoRA fine-tuning, we adopt a standard parameter-efficient configuration. Specifically, we set the LoRA rank to $r=16$, scaling factor $\alpha =16$, and apply a dropout rate of $0.05$ to the LoRA layers. The model is optimized using AdamW with a learning rate of 2e-4 and trained for multiple epochs. 

Table~\ref{tab:peft_vs_icl} summarizes the performance comparison on binary classification. Overall, the results reveal a clear data-efficiency advantage of ICL. Notably, ICL with 200 demonstrations achieves an F1-score of 0.943 and an accuracy of 0.944, closely matching the performance of a LoRA-tuned model trained on 4,480 labeled examples. This corresponds to a reduction of more than 22$\times$ in the number of task-specific supervised samples. Moreover, ICL consistently exhibits a lower false positive rate (FPR), indicating more precise and conservative decision boundaries.

\begin{table}[htbp]
    \centering
    \caption{\textbf{Data Efficiency Comparison.} Performance benchmarking of ICL ($k$-shot) versus LoRA fine-tuning (trained on $K$ samples) for binary classification.}
    \label{tab:peft_vs_icl}
    \begin{tabular}{cc@{\hspace{0.5em}}c@{\hspace{0.5em}}cccccc} 
        \toprule
        \textbf{Method} & \textbf{K/k} & \textbf{Epochs} & \textbf{Prec.} & \textbf{Rec.} & \textbf{F1} & \textbf{FPR} & \textbf{Acc.} \\
        \midrule
        \multirow{15}{*}{PEFT} 
        & \multirow{3}{*}{280} & 1 & 0.5632 & 0.9900 & 0.7166 & 0.7142 & 0.6220 \\
        & & 2 & 0.6035 & 0.9918 & 0.7473 & 0.6148 & 0.6738 \\
        & & \textbf{3} & \textbf{0.7692} & \textbf{0.9750} & \textbf{0.8538} & \textbf{0.3007} & \textbf{0.8310} \\
        \cmidrule(lr){2-8}  
        & \multirow{3}{*}{560} & 1 & 0.5627 & 0.9994 & 0.7176 & 0.7283 & 0.6180 \\
        & & 2 & 0.7385 & 0.9899 & 0.8433 & 0.3393 & 0.8187 \\
        & & \textbf{3} & \textbf{0.7724} & \textbf{0.9893} & \textbf{0.8665} & \textbf{0.2702} & \textbf{0.8535} \\
        \cmidrule(lr){2-8}  
        & \multirow{3}{*}{1120} & 1 & 0.8298 & 0.9834 & 0.8998 & 0.1872 & 0.8947 \\
        & & 2 & 0.8477 & 0.9830 & 0.9087 & 0.1689 & 0.9042 \\
        & & \textbf{3} & \textbf{0.8580} & \textbf{0.9871} & \textbf{0.9167} & \textbf{0.1565} & \textbf{0.9126} \\
        \cmidrule(lr){2-8}  
        & \multirow{3}{*}{2240} & \textbf{1} & \textbf{0.8817} & \textbf{0.9854} & \textbf{0.9304} & \textbf{0.1266} & \textbf{0.9279} \\
        & & 2 & 0.8666 & 0.9862 & 0.9218 & 0.1468 & 0.9178 \\
        & & 3 & 0.8743 & 0.9848 & 0.9262 & 0.1322 & 0.9242 \\
        \cmidrule(lr){2-8}  
        & \multirow{3}{*}{4480} & 1 & 0.9105 & 0.9886 & 0.9479 & 0.0914 & 0.9474 \\
        & & \textbf{2} & \textbf{0.9174} & \textbf{0.9884} & \textbf{0.9516} & \textbf{0.0824} & \textbf{0.9516} \\
        & & 3 & 0.8546 & 0.9924 & 0.9183 & 0.1561 & 0.9152 \\
        \midrule
        \multirow{2}{*}{ICL} & 128 & -- & 0.9232 & 0.9554 & 0.9390 & 0.0732 & 0.9405 \\
        & 200 & -- & 0.9382 & 0.9474 & 0.9428 & 0.0584 & 0.9444 \\
        \bottomrule
    \end{tabular}
\end{table}

The comparison further highlights the substantial overhead associated with fine-tuning in low-data regimes. When trained on small datasets ($K=280$ or $560$), LoRA models struggle to achieve competitive performance, with F1-scores remaining around 0.86 even under carefully selected epoch settings. In contrast, as shown in \S\ref{sec:effective_shots}, ICL reaches a similar performance level using only 32–64 demonstrations, underscoring its immediate applicability in data-scarce settings where collecting labeled samples is costly or impractical.

Another critical distinction emerges in terms of training stability. The performance of LoRA is highly sensitive to the choice of training epochs, particularly when the amount of training data is limited. For larger datasets, we observe clear signs of overfitting, such as the performance degradation at epoch 3 for $K=4480$. ICL, by design, avoids this issue entirely: it requires no iterative optimization or hyperparameter tuning, and its behavior is directly controlled by the number and composition of demonstrations. This property makes ICL substantially more predictable and robust in deployment scenarios.

\begin{figure}[htbp]
\centering
\includegraphics[width=0.36\textwidth]{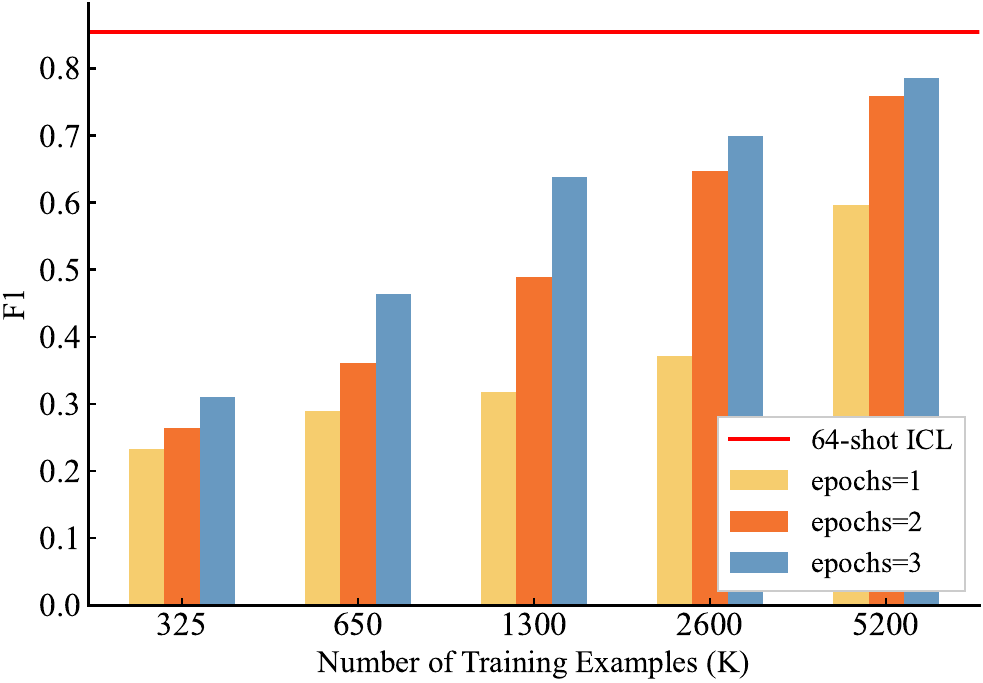}
\caption{\textbf{Multiclass Efficiency Benchmark.} Comparing 64-shot ICL against LoRA fine-tuning with increasing training sample sizes ($K$) and epochs.}
\label{fig:peft_vs_icl_multi}
\end{figure}

The advantage of ICL becomes even more pronounced in complex multiclass classification tasks. As shown in Fig.~\ref{fig:peft_vs_icl_multi}, a LoRA model fine-tuned on the full multiclass training set ($K=5,200$) is outperformed by an ICL configuration using only 64 demonstrations. This result suggests that for fine-grained categorization tasks, where learning nuanced and class-specific decision boundaries is critical, the dynamic and query-adaptive reasoning enabled by ICL is inherently more effective than the static parameter updates produced by fine-tuning on limited data. By presenting carefully selected, semantically aligned examples at inference time, ICL can better navigate complex semantic spaces and capture subtle distinctions among illicit categories.

\subsection{Robustness to Prompt Ordering and Positional Bias}
The data efficiency advantage makes ICL a powerful alternative to fine-tuning. However, for reliable deployment in production systems, a method must also exhibit robustness, i.e., its performance should not be brittle to innocuous variations in input formatting. A particular concern for ICL is its potential sensitivity to the order of elements within the dynamic prompt. This section systematically evaluates the robustness of ICL against two such variations: the order of demonstration examples and the order of labels in the demonstrations.

\noindent\textbf{Sensitivity to Demonstration Ordering.}
In a dynamically constructed few-shot prompt, the sequence of demonstration examples is typically arbitrary. To assess whether this order influences model predictions, we conducted experiments where we held the set of $k$ demonstrations constant but randomly shuffled their sequence multiple times, measuring the variation in performance (standard deviation of F1-score) for the Mistral model on both binary and multiclass classification tasks.

\begin{figure}[htbp]
\centering
\includegraphics[width=0.36\textwidth]{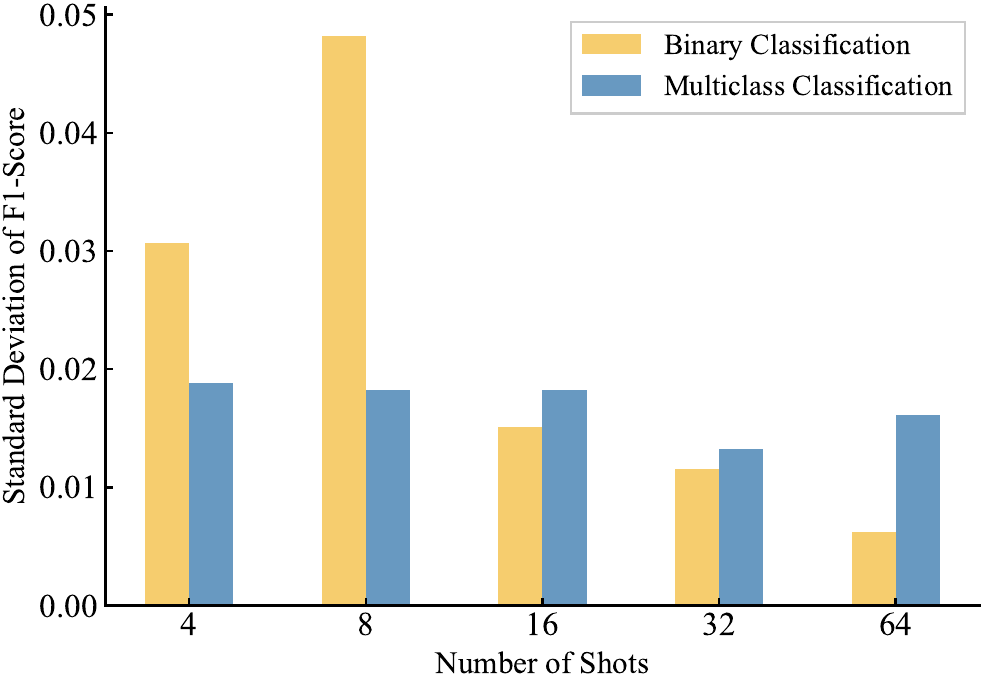}
\caption{\textbf{Robustness to Demonstration Ordering.} Standard deviation of F1-scores across random permutations of the same demonstration set, illustrating stability improvements as shot count increases.}
\label{fig:shotorder}
\end{figure}

Fig.~\ref{fig:shotorder} reports the standard deviation of F1-score across multiple random permutations of the same demonstration set for different numbers of shots. Overall, the results indicate that demonstration ordering has a measurable but diminishing impact on model performance as the number of shots increases.

For binary classification, the sensitivity to demonstration order is most pronounced in low-shot settings. With only 4 and 8 demonstrations, the standard deviation reaches 0.0307 and 0.0482, indicating substantial variability induced solely by reordering examples. However, this sensitivity decreases rapidly as more demonstrations are provided. At 16 shots, the standard deviation drops to 0.0151, and further declines to 0.0062 at 64 shots, respectively. This trend suggests that a larger demonstration set provides a more stable and redundant contextual signal, mitigating the influence of individual example positions.

In contrast, multiclass classification exhibits a more moderate and consistent sensitivity across shot counts. The standard deviation remains around 0.018 for 4 to 16 shots and decreases to 0.0132 at 32 shots, before slightly increasing to 0.0161 at 64 shots. Compared to the binary setting, this relatively stable pattern implies that multiclass performance variability is less dominated by the ordering of individual demonstrations and more influenced by the inherent complexity of the task and class diversity.

Taken together, these findings suggest that while demonstration ordering can introduce non-negligible variance in few-shot ICL, particularly in low-shot binary classification, its impact becomes marginal as the number of demonstrations increases. From a practical perspective, this implies that for moderate to large $k$, ICL systems can safely ignore demonstration order without sacrificing robustness, whereas in extremely low-shot regimes, controlling or averaging over multiple prompt orderings may be beneficial.

\noindent\textbf{Sensitivity to Label Ordering.}
Beyond the ordering of demonstrations themselves, we examine a more subtle source of bias: whether the model’s predictions are influenced by the positional distribution of labels within the prompt, independent of their semantic relevance. Specifically, we investigate whether the model may implicitly favor labels that appear more frequently or more recently in the demonstration sequence, thereby introducing systematic distortions in its decision behavior. 

\begin{figure}[htbp]
    \centering
    \subfloat[Accuracy.\label{fig:labelorder_acc}]{%
        \includegraphics[width=0.24\textwidth]{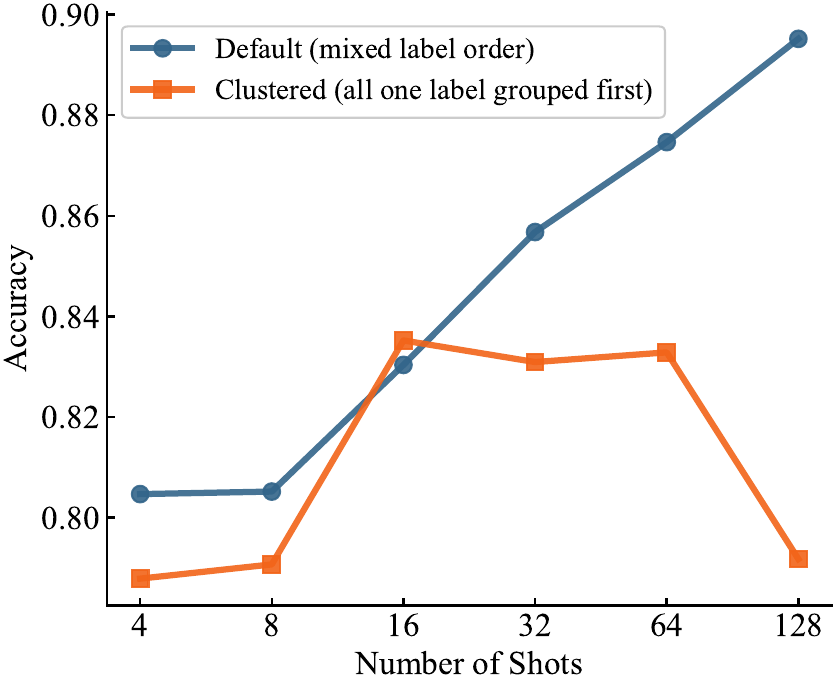}%
    }
    \hfill
    \subfloat[False positive rate.\label{fig:labelorder_fpr}]{%
        \includegraphics[width=0.24\textwidth]{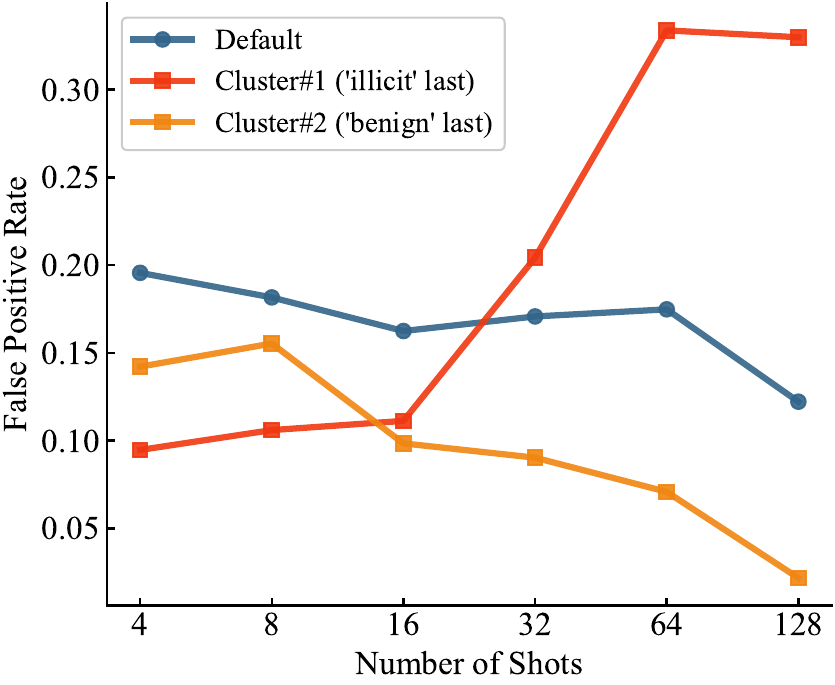}%
    }
    \caption{\textbf{Analysis of Positional Label Bias.} (a) Accuracy degradation under clustered label arrangements compared to mixed order. (b) Recency Bias: The False Positive Rate significantly fluctuates depending on whether the label nearest to the query is 'illicit' or 'benign'.}
    \label{fig:labelorder}
\end{figure}

We manipulate the global arrangement of class labels across the entire demonstration set while keeping the underlying examples unchanged. Concretely, all demonstrations belonging to the same class are grouped together and placed either at the beginning or the end of the prompt. As shown in Fig.~\ref{fig:labelorder_acc}, this global reordering leads to a noticeable degradation in accuracy when the number of shots becomes large ($k \geq 32$), compared to the default mixed-label order. This effect suggests that even when demonstrations are relevant, a coherent and imbalanced label signal can bias the model’s implicit decision boundary, particularly when accumulated over many examples.

More strikingly, we observe a strong positional or \emph{recency} bias in the model’s predictions. As illustrated in Fig.~\ref{fig:labelorder_fpr}, the label that appears most frequently in close proximity to the query exerts a disproportionate influence on the final output. When \textit{illicit} demonstrations are placed near the end of the prompt, the model becomes more inclined to predict the illicit label, resulting in a substantially higher false positive rate. Conversely, positioning \textit{benign} demonstrations closest to the query significantly suppresses false positives, albeit at the risk of reduced recall for illicit cases. 

This phenomenon, which is also observed in a controlled multiclass setting, indicates that the model assigns greater weight to local contextual evidence near the query than to globally aggregated information. Such behavior suggests that ICL-based classifiers do not rely solely on abstract class semantics but are sensitive to positional cues within the prompt, which can systematically skew predictions if not carefully controlled.

\subsection{Contextual Information Retrieval Capabilities}
Building upon the demonstrated advantages of ICL in terms of data efficiency and robustness, we now turn to a more mechanistic question: how effectively can an LLM identify and exploit relevant information within a complex, and potentially noisy, in-context prompt?

To this end, we design a \emph{Needle-in-a-Haystack} experiment that explicitly probes the model’s ability to retrieve a single, highly relevant demonstration from a prompt dominated by irrelevant context. This setting allows us to disentangle genuine in-context retrieval from superficial pattern matching or majority-label biases.

In the experiment setup, for each target query in the test set, we construct a prompt containing $n$ demonstrations, where $n \in \{2, 4, 8, 16, 32, 64, 128\}$. Exactly one demonstration is an identical copy of the target query paired with its ground-truth label (the \emph{needle}), while the remaining $n-1$ demonstrations are randomly sampled from the training set and are semantically unrelated to the query (the \emph{haystack}). The position of the needle within the demonstration sequence is uniformly randomized to avoid positional artifacts. The model is then asked to classify the target query, as in the standard ICL setting.
If the model perfectly retrieves and copies the label from the needle, accuracy should remain 100\% regardless of $n$. Any drop in accuracy thus directly measures the model’s failure to isolate the relevant signal amidst increasing noise.

\begin{figure}[htbp]
\centering
\includegraphics[width=0.36\textwidth]{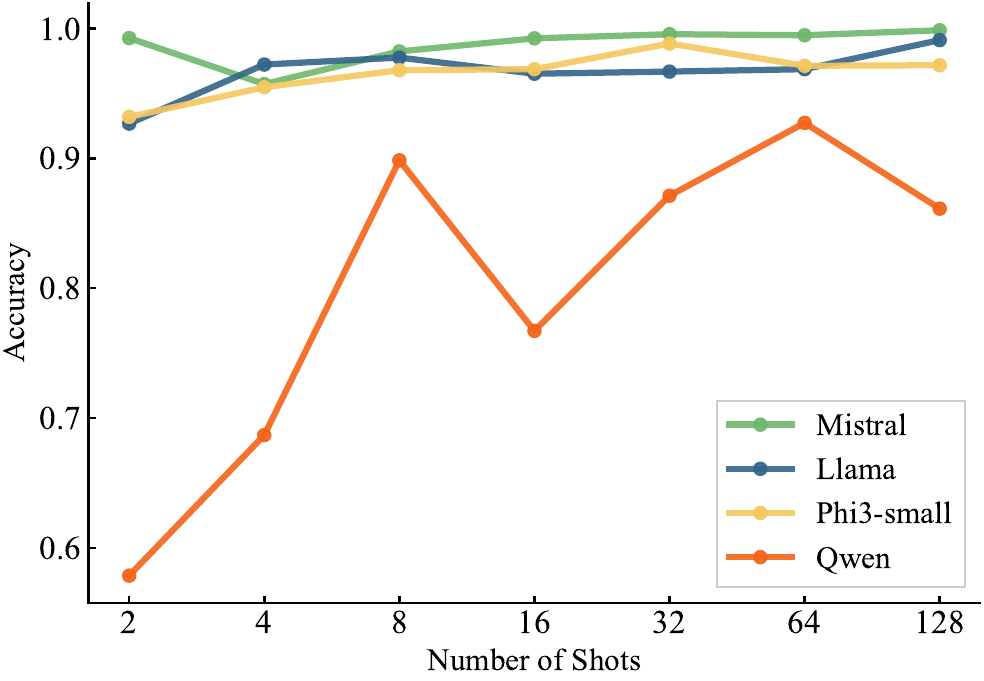}
\caption{\textbf{Contextual Retrieval Capability (Needle-in-a-Haystack)}. Accuracy of models in retrieving a single relevant demonstration amidst an increasing number of irrelevant "haystack" examples.}
\label{fig:haystack}
\end{figure}

The results across four model families are shown in Fig.~\ref{fig:haystack}. Mistral, Llama, and Phi3-small show strong noise resilience, maintaining near-perfect accuracy even with 128 demonstrations. This indicates a robust capacity to attend to the relevant signal despite substantial distraction.
In contrast, Qwen shows pronounced sensitivity to the increasing size of the haystack. While its accuracy initially rises with more demonstrations, it declines noticeably at larger $n$ values, suggesting difficulty in consistently identifying the needle when overwhelmed by irrelevant content.

These findings provide direct empirical evidence that ICL performance is not solely a function of demonstration quantity, but critically depends on the model’s ability to selectively retrieve and prioritize relevant input information. Models with stronger contextual retrieval capabilities are better equipped to scale ICL to longer prompts without suffering from information dilution, while models that struggle in this setting are more susceptible to noise.

Overall, the Needle-in-a-Haystack experiment highlights a fundamental but often overlooked aspect of ICL: its reliance on implicit, model-dependent retrieval mechanisms over the input context. This insight underscores the importance of model selection when deploying ICL in real-world applications with long or noisy contexts.

\section{Deployment in Open-World Scenarios}
\label{sec:apply}
The analyses demonstrated that ICL is not only data-efficient but, when utilizing capable models, exhibits strong resilience to noise and irrelevant context. Leveraging these robust retrieval capabilities, we now advance to the most challenging phase: deploying ICL in open-world scenarios. In this section, we evaluate whether ICL can translate its mechanistic strengths into practical utility, specifically testing its ability to generalize to unseen categories, autonomously discover novel threats, and operate across heterogeneous platforms without adaptation.

\subsection{Generalization to Unseen Illicit Categories}
A practical content moderation system must be resilient not only to known forms of illicit activity, but also to novel and previously unseen promotion categories that continuously emerge in online ecosystems. One of the key promises of ICL is its ability to extrapolate from a limited set of demonstrations to new but semantically related tasks without parameter updates. To evaluate this capability rigorously, we design a controlled experiment simulating a deployment scenario where the system encounters content from a completely unseen illicit category.

We frame this task as a binary classification problem with labels \textit{benign} and \textit{illicit}, using the instruction-tuned Mistral model. Each prompt contains 64 demonstrations, evenly split between benign and illicit examples, retrieved randomly to avoid category-specific bias from semantic search.

For each of the 12 illicit categories, we construct three evaluation settings:
(i) \textit{Zero-shot}: no demonstrations are provided;
(ii) \textit{Excluded}: demonstrations include benign samples and illicit samples from all categories except the target;
(iii) \textit{Included}: demonstrations additionally contain a few samples from the target category.
In all cases, the test set consists exclusively of samples from the target category, ensuring that accuracy directly reflects the model’s ability to generalize to that category (with precision fixed at 1).

Results are summarized in Table~\ref{tab:unseen_generalization}. Several consistent patterns emerge. First, using ICL demonstrations that exclude the target category yields performance improvements over the zero-shot baseline for the majority of categories. In more than half of the cases, the performance drop incurred by excluding the target category is less than 6\% relative to the category-included setting, indicating that ICL can effectively generalize from other illicit categories to unseen ones.

Second, while including demonstrations from the target category unsurprisingly produces the best overall performance, the gains are often modest compared to the excluded setting. This suggests that much of the discriminative capability arises from shared semantic and structural characteristics across illicit promotion types, rather than from category-specific memorization.

Notably, categories such as \textit{fraud}, \textit{data-theft}, and \textit{counterfeit} exhibit substantial improvements over zero-shot inference even when excluded from the demonstrations, highlighting ICL’s strong cross-category generalization. In contrast, a small subset of categories (e.g., \textit{advertisement} and \textit{surrogacy}) show larger gaps between included and excluded settings, reflecting higher semantic divergence from other illicit categories and thus a greater reliance on category-specific cues.

These findings provide empirical evidence that ICL can generalize beyond seen categories by leveraging high-level shared patterns. Even without target-category demonstrations, the model often infers malicious intent from contextual similarities learned from other categories. This capability is particularly valuable in real-world deployment scenarios, where new forms of illicit promotion frequently appear before labeled data can be collected.

Overall, this experiment shows that ICL exhibits meaningful extrapolative generalization across illicit categories, supporting its suitability as a flexible and adaptive moderation approach.

\begin{table}[htbp]
\centering
\caption{\textbf{Generalization performance of ICL on unseen illicit categories.}
Accuracy is reported under three settings: zero-shot, category-included demonstrations, and category-excluded demonstrations. The testing set of each row contains only samples from the target category.}
\label{tab:unseen_generalization}
\begin{tabular}{l@{\hspace{0.5em}}c@{\hspace{0.5em}}c@{\hspace{0.5em}}c@{\hspace{0.5em}}c@{\hspace{0.5em}}c}
\toprule
\multirow{2}{*}{\textbf{Category}} 
& \multirow{2}{*}{\textbf{Zero-shot}} 
& \multirow{2}{*}{\textbf{Included}} 
& \multirow{2}{*}{\textbf{Excluded}} 
& \textbf{Gain over} 
& \textbf{Gap vs.} \\
& & & & {\textbf{Zero-shot}}
& \textbf{Included} \\
\midrule
Hacking         & 0.9011 & 0.9457 & 0.9550 & +0.0539 & +0.0093 \\
Others          & 0.8861 & 0.9080 & 0.9080 & +0.0218 &  0.0000 \\
Drug            & 0.9103 & 0.9733 & 0.9558 & +0.0454 & -0.0176 \\
Counterfeit     & 0.6828 & 0.8816 & 0.8571 & +0.1743 & -0.0244 \\
Data-theft      & 0.6916 & 0.9574 & 0.9225 & +0.2309 & -0.0349 \\
Porn            & 0.8497 & 0.8643 & 0.8221 & -0.0276 & -0.0421 \\
Fraud           & 0.5124 & 0.9378 & 0.8780 & +0.3657 & -0.0597 \\
Benign          & 0.8789 & 0.8426 & 0.7778 & -0.1011 & -0.0648 \\
Gambling        & 0.5988 & 0.8035 & 0.6711 & +0.0723 & -0.1324 \\
Weapon          & 0.6600 & 0.9393 & 0.7934 & +0.1334 & -0.1459 \\
Money-laundering& 0.8485 & 0.9118 & 0.7595 & -0.0890 & -0.1523 \\
Surrogacy       & 0.7500 & 0.9880 & 0.7640 & +0.0140 & -0.2240 \\
Advertisement   & 0.7263 & 0.9177 & 0.6266 & -0.0998 & -0.2911 \\
\bottomrule
\end{tabular}
\end{table}

\subsection{Autonomous Discovery of Emerging Threats}
\noindent\textbf{Motivation.}
A persistent challenge in online content moderation is the rapid emergence of novel illicit goods and services. In practice, any fixed classification taxonomy inevitably results in detection gaps, leaving novel threats uncaught. Therefore, enabling a system to autonomously discover and categorize previously unknown forms of illicit promotion from unlabeled data is crucial for proactive defense.

In principle, this problem could be addressed through periodic taxonomy updates driven by manual inspection. However, such expert-driven processes are slow, labor-intensive, and difficult to scale in the face of high-throughput online platforms. A natural automated alternative is unsupervised clustering over textual representations. Unfortunately, our empirical evaluation shows that conventional clustering methods (e.g., DBSCAN on word-level or sentence-level embeddings) are ill-suited for this domain. These approaches tend to produce overly coarse semantic groupings that lack discriminative granularity. For example, distinct illicit activities such as \textit{hacking}, \textit{counterfeit trading}, and \textit{fraud} are merged into a single cluster, while categories like \textit{surrogacy}, \textit{gambling}, and \textit{pornography} are conflated into another. This failure arises because such methods rely on general lexical or distributional similarity and fail to capture the nuanced distinctions between fine-grained illicit categories. The resulting clusters are thus too vague to support actionable threat intelligence.

Therefore, these limitations highlight the need for an alternative approach that preserves the scalability of automation while incorporating the contextual and domain-aware reasoning typically associated with human analysis.

\noindent\textbf{Methodology.}
To address this challenge, we propose a two-stage pipeline that leverages the generative and reasoning capabilities of LLMs to perform both category discovery and conceptual consolidation.

\noindent\textit{Stage 1: ICL-based Category Annotation.}
Rather than forcing content into a fixed label set, we prompt an LLM to operate as a human-like content analyst. Given an unlabeled promotional text, the model is asked to generate a concise, free-form category name that describes the core illicit activity being promoted. This step is implemented using a few-shot ICL configuration with the Mistral model. We explore two prompt designs to examine the impact of prior structural constraints:
\begin{itemize}
    \item \textbf{Anchored Prompting.} The prompt provides a list of known illicit categories as references, while explicitly allowing the model to propose a new category name if none of the existing ones are suitable.
    \item \textbf{Open-Ended Prompting.} The prompt provides no predefined categories and simply instructs the model to propose a category name that best captures the illicit activity described in the text. This design minimizes anchoring bias and encourages unconstrained conceptualization.
\end{itemize}

\noindent\textit{Stage 2: LLM-powered Conceptual Clustering.}
The first stage produces a large set of heterogeneous, free-form category strings (e.g., \textit{luxury-goods-counterfeit}, \textit{live-streaming-sex}). To transform this unstructured output into a usable taxonomy, we perform a second LLM invocation. The model is provided with the complete list of generated category names and instructed to group them based on their underlying illegal activities, assign a concise and representative cluster name to each group, and produce a brief semantic summary. This step leverages the LLM’s ability to reason over conceptual similarity rather than relying on superficial lexical overlap, enabling coherent consolidation of semantically related but lexically diverse labels.

\noindent\textbf{Results.}
The impact of prompt design is immediately reflected in the diversity of generated category names. The anchored prompting strategy yields 197 unique category labels, whereas the open-ended strategy produces 2,909 unique labels, demonstrating a substantially richer exploration of the category space when prior constraints are removed.

More importantly, qualitative analysis of the final clustered outputs reveals the pipeline’s capacity for genuine category discovery. The anchored pipeline, while constrained by existing knowledge, identifies two coherent novel clusters absent from the original taxonomy. In contrast, the open-ended pipeline discovers eight distinct new illicit categories, including emerging or niche threats such as \textit{usury}, \textit{illegal immigration}, and \textit{software piracy}. These categories are not only semantically coherent but also practically meaningful, aligning well with real-world moderation concerns.

Overall, this experiment demonstrates that ICL can be effectively repurposed from a static classification mechanism into a dynamic discovery engine. By decoupling label generation from predefined schemas and subsequently imposing structure through LLM-based conceptual clustering, our two-stage design enables open-world exploration while maintaining organizational clarity. This approach shifts the role of human analysts from exhaustive manual discovery to efficient validation and oversight. As a result, it substantially accelerates the threat intelligence lifecycle and moves content moderation systems closer to truly adaptive, future-proof operation.

\subsection{Cross-Platform Evaluation in the Wild}

We evaluate the configured ICL under real operational conditions, using large-scale, unlabeled data collected from heterogeneous online platforms. This setting represents the most challenging scenario in our study: highly imbalanced class distributions, platform-specific linguistic noise, and the absence of ground-truth supervision. Our goal is not merely to benchmark raw accuracy, but to assess whether ICL offers unique practical advantages that are difficult to replicate with conventional, platform-specific detectors.

\noindent\textbf{Experimental Setup.}
We collected 100,000 unlabeled samples each from search engine results and Twitter tweets. The optimized ICL pipeline, utilizing the Mistral model with 32-shot semantic retrieval, was applied without platform-specific modifications. Performance was evaluated through manual auditing based on standard industry practices, with 1,000 randomly selected predictions reviewed for each platform. As baselines, we compared ICL against two state-of-the-art, platform-specific detectors: (i) a high-efficiency Random Forest classifier for Search Engine data, and (ii) a fine-tuned, transformer-based multilingual model for Twitter data.

\noindent\textbf{Cross-Platform Generalization on Mixed Data.}
To directly assess cross-platform generalization, we construct a curated, balanced test set that equally mixes samples from search engine and Twitter sources. 
As shown in Table~\ref{tab:performance_mixed_set}, ICL achieves the best overall performance across all evaluation metrics, with an accuracy of 92.59\% and an F1-score of 92.31\%. In contrast, both platform-specific baselines suffer substantial degradation when applied outside their native domains. The search-engine Random Forest classifier, optimized for efficiency and lexical patterns in query data, performs poorly on Twitter content, resulting in a sharp drop in both recall and overall accuracy. Similarly, the Twitter fine-tuned transformer model, while achieving high precision, exhibits a pronounced recall loss on search engine samples, reflecting its limited exposure to non-social-media distributions during training.

Importantly, ICL’s superior performance is achieved without any platform-aware adaptation or retraining. A single, unified model and prompting strategy is applied identically across both data sources. This contrasts fundamentally with the platform-specific detectors, whose performance is tightly coupled to their training distributions and degrades under distribution shift. These results demonstrate that ICL functions as a genuinely platform-agnostic classifier, capable of leveraging semantic reasoning over in-context demonstrations to bridge heterogeneous data sources.

\begin{table}[htbp]
\centering
\begin{threeparttable}
\caption{\textbf{Cross-Platform Generalization.} Performance comparison between ICL and platform-specific baselines (Random Forest and Fine-tuned Transformer) on a balanced, mixed-source test set.}
\label{tab:performance_mixed_set}
\begin{tabular}{lccccc}
\toprule
\textbf{Classifier} & \textbf{Accuracy} & \textbf{Precision} & \textbf{Recall} & \textbf{F1-score} & \textbf{FPR} \\
\midrule
ICL (Ours) & 0.9259 & 0.9111 & 0.9354 & 0.9231 & 0.0826 \\
Random Forest\tnote{1} & 0.7330 & 0.7202 & 0.7162 & 0.7182 & 0.2517 \\
Transformer\tnote{2} & 0.8339 & 0.8728 & 0.7613 & 0.8133 & 0.1003 \\
\bottomrule
\end{tabular}
\begin{tablenotes}
    \footnotesize
    \item[1] This is search engine-specific.
    \item[2] This is Twitter-specific.
\end{tablenotes}
\end{threeparttable}
\end{table}

\noindent\textbf{Behavior on Wild Data.}
When deployed on fully unlabeled wild data, ICL demonstrates a distinct operational profile that differs markedly from platform-specific detectors. On search engine data, ICL achieves an F1-score of 83.89\%, closely approaching the performance of the specialized search engine classifier, despite operating without any platform-specific adaptation or retraining.

On Twitter data, ICL exhibits a high-recall but lower-precision behavior, achieving a recall of 92.41\% and a precision of 68.73\%, which leads to an elevated false positive rate compared to the Twitter-specific model. This performance pattern does not indicate a failure of the approach, but rather reflects a fundamental difference in design philosophy. Platform-specific detectors are typically optimized for high-precision filtering under relatively stable, closed-world distributions. In contrast, ICL prioritizes semantic coverage and adaptability, aggressively surfacing content that weakly matches known illicit patterns but may correspond to emerging, obfuscated, or borderline promotional behaviors.

To further investigate this phenomenon, we conduct a comparative analysis of samples uniquely flagged by each system. Manual inspection reveals that 61.8\% of the samples labeled as illicit exclusively by ICL correspond to either clearly illicit or borderline promotional content, often involving implicit language, novel phrasing, or cross-domain references. In comparison, among samples flagged only by the Twitter-specific classifier, this proportion drops to 45.7\%, with the remaining cases largely attributable to platform-specific artifacts or overly narrow lexical cues.

Taken together, these results position ICL as a complementary paradigm rather than a replacement for traditional classifiers. Its strengths lie in cross-platform generalization, zero-adaptation deployment, and the discovery of emerging illicit behaviors, and these capabilities are inherently difficult to achieve with fine-tuned, platform-specific models. In practice, the most effective deployment strategy is a hybrid system in which ICL operates as a flexible, high-recall discovery engine, continuously feeding novel patterns and labeled data to efficient downstream classifiers. Such a design enables adaptive, future-proof content moderation in the face of rapidly evolving online threats.

\section{Discussion}
\label{sec:discussion}
\subsection{Implications for Practical Content Moderation Systems}
While ICL demonstrates strong data efficiency and cross-platform generalization, its deployment entails trade-offs, including higher inference latency and lower precision on noisy data. Consequently, we position ICL not as a standalone replacement for specialized classifiers, but as the core of a tiered, hybrid architecture.

In scenarios where labeled data are scarce, costly, or delayed, ICL can function effectively as a primary detector. Our analysis confirms that ICL achieves competitive performance using up to 22 times fewer labeled samples than fine‑tuned models. This allows for immediate operational coverage while structured data‑collection pipelines are being established, reducing dependence on large‑scale annotation efforts and accelerating response to emerging promotion tactics.

For platforms with mature moderation systems, ICL is best deployed in parallel with traditional classifiers as a high‑recall discovery funnel. Its strength in generalizing from limited demonstrations enables it to flag ambiguous, novel, or intentionally obfuscated content that may evade static rule‑based or fine‑tuned models. These high-value samples can then be reviewed by human experts to generate labeled data for updating lightweight downstream models, creating a virtuous cycle of model improvement.

In summary, this synergistic architecture combines the agility and generalizing power of ICL with the efficiency and precision of specialized classifiers, establishing a more robust and adaptive defense against the continuously evolving challenge of illicit online promotion.

\subsection{Limitations and Future Work}
\label{sec:limitations}


\noindent\textbf{Restriction to Textual Modality.}
Our evaluation focuses primarily on text-based illicit promotion. In real-world ecosystems, illicit promotion increasingly adopts multimodal strategies, combining text with images, videos, URLs, and encoded symbols to evade detection. While ICL exhibits strong adaptability in the textual domain, its effectiveness for multimodal illicit promotion detection remains unexplored. Extending the proposed framework to multimodal large language models and integrating visual and structural signals constitutes an important avenue for future work.

\noindent\textbf{Dependence on Demonstration Quality and Diversity.}
Although ICL shows strong generalization to unseen illicit categories, its performance remains dependent on the quality and diversity of the demonstration pool. When demonstrations are biased, outdated, or semantically narrow, the model may inherit these limitations and exhibit reduced robustness. Future work could investigate adaptive demonstration management strategies, such as continuously refreshing the demonstration repository through active learning or uncertainty-driven sampling, in order to maintain coverage over emerging threat patterns.

\noindent\textbf{Inference Latency and Cost.} While ICL eliminates training costs, utilizing long-context prompts incurs significant computational overhead during inference compared to lightweight fine-tuned classifiers. This latency trade-off limits the direct applicability of high-shot ICL for real-time, high-throughput filtering. Future work should investigate prompt compression techniques or knowledge distillation, where ICL models serve as "teachers" to generate rationale-augmented training data for smaller, latency-efficient student models.

\subsection{Data and Code Release}
\label{sec:data_release}

To support reproducibility and facilitate future research, we plan to release the data and code associated with this study, subject to ethical and legal constraints.
Specifically, we will release:
(i) the full implementation of our ICL pipeline, including prompt templates, semantic retrieval strategies, and evaluation scripts; and
(ii) anonymized and de-identified versions of the curated benchmark datasets used in controlled experiments.
All released resources are available and will be continuously updated
on \url{https://github.com/ChaseSecurity/illicit-icl}. We hope that these resources will enable the community to replicate our findings, explore alternative prompt designs, and extend ICL-based approaches to new domains of online security.

\section{Conclusion}
\label{sec:conclusion}
In this paper, we investigated in-context learning as a practical and adaptive approach for illicit promotion detection across online platforms. Through extensive experiments, we showed that ICL exhibits robust cross-platform generalization, meaningful resilience to unseen categories, and a unique capacity for autonomous threat discovery. These capabilities that are difficult to achieve with traditional fine-tuning-based systems.
At the same time, our analysis highlights important limitations, including sensitivity to prompt construction and reduced precision on highly noisy data. These findings underscore that ICL should not be viewed as a universal replacement for specialized classifiers, but rather as a complementary paradigm with distinct strengths.

By reframing illicit content detection as an open-world, evolving problem, our work demonstrates that ICL enables a shift from static classification toward adaptive, discovery-driven moderation. We believe this perspective opens new directions for future research, including more principled prompt optimization, integration with retrieval systems, and tighter coupling between ICL-driven discovery and downstream learning.

        \bibliographystyle{plain}
        \bibliography{ref}
        
        \appendices
        \label{sec:appendix}
        \section{The Selection of Model.}
\label{app:mistral}
\begin{figure}[h]
    \centering
    \subfloat[Binary classification.\label{fig:binary_semantic_comparison}]{%
        \includegraphics[width=0.24\textwidth]{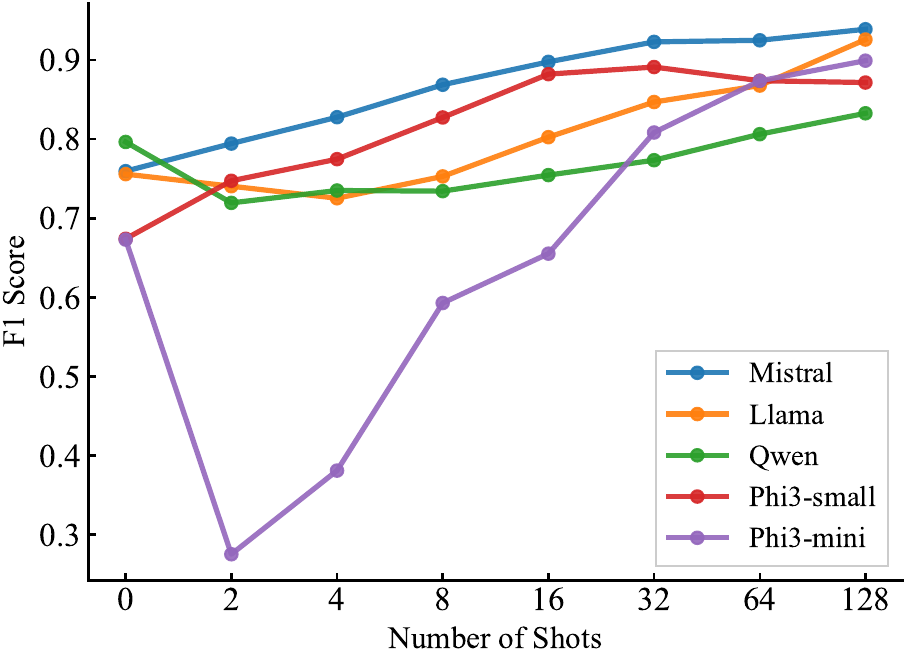}%
    }
    \hfill
    \subfloat[Multiclass classification.\label{fig:multi_semantic_comparison}]{%
        \includegraphics[width=0.24\textwidth]{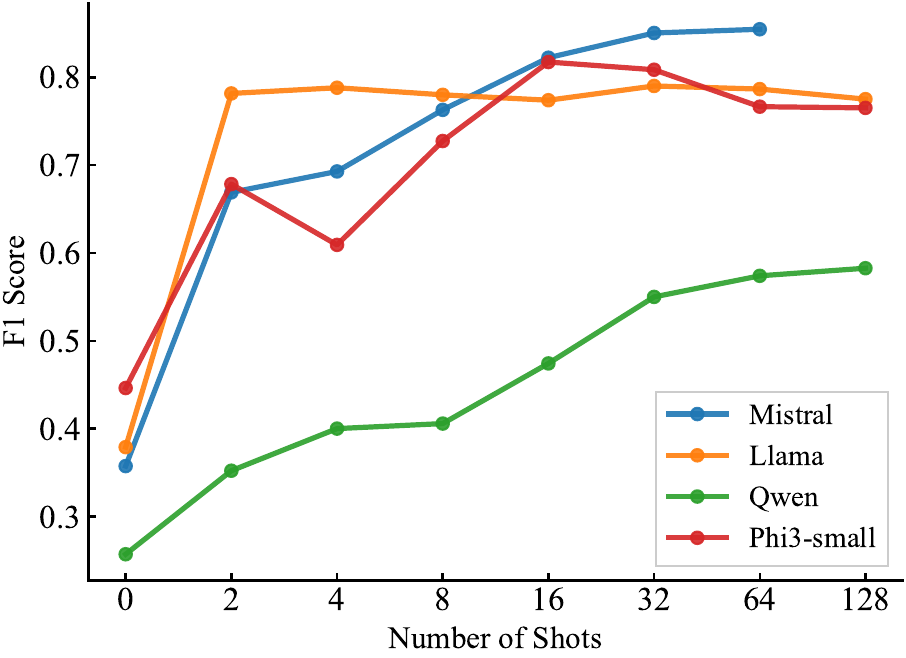}%
    }
    \caption{\textbf{Model Selection Basis.} Performance comparison of all candidate models, justifying the selection of Mistral as the primary backbone for subsequent experiments due to its superior multiclass performance.}
    \label{fig:semantic_comparison}
\end{figure}

We compared the performance of different models under semantic-based retrieval and found that Mistral performed the best across all tasks, as shown in Figure~\ref{fig:semantic_comparison}. Therefore, we selected Mistral as the baseline model for all subsequent experiments.

\section{Example Prompts.}
\label{app:prompt}

\vspace{1em}
\begin{promptbox}[Binary Classification Task Prompt Template]
You will act as a content detector to decide if a given text is illicit promotion, i.e., involving the promotion of illicit goods and services.

If it is, label it as 'illicit', otherwise, label it as 'benign'.

First, you will be given a set of message examples along with their labels. Then, you will answer the category of the last message from the following options:

- 'benign': Content that does not contain illicit promotion

- 'illicit': Content that contain illicit promotion

Return only the name of the category, and nothing else.

==

Query: \{demonstration\_text\}

Answer: \{demonstration\_label\}

==

\{...\}

==

Query: \{query\_text\}

Answer: 
\end{promptbox}

\begin{promptbox}[Multiclass Classification Task Prompt Template]
You will act as a content detector.

First, you will be given a set of message examples along with their labels. Then, you will answer the category of the last message from the following options:

- 'benign': Content that does not contain illicit promotion

- 'porn': Content that is sexually explicit or pornographic in nature

- 'surrogacy': Content related to surrogate mothers and gestational surrogacy

- 'gambling': Content related to gambling or betting activities

- 'drug': Content related to illegal drug use, sales, or promotions

- 'data-theft': Content involving the theft or illegal use of data, identity theft, or similar activities

- 'money-laundry': Content involving the promotion or recruitment for money laundering activities

- 'counterfeit': Content related to fake goods, forged certificates or false accounts

- 'advertisement': Content related to illegal marketing and black hat SEO

- 'weapon': Content related to weapons, including sales, manufacturing, or usage

- 'fraud': Content related to fraudulent activities and scams

- 'hacking': Content related to hacking, cybersecurity threats, or development of unlawful programs

- 'others': Content that does not fit into any of the above categories

Return only the name of the category, and nothing else.

==

Query: \{demonstration\_text\}

Answer: \{demonstration\_label\}

==

\{...\}

==

Query: \{query\_text\}

Answer: 
\end{promptbox}
            
    \end{CJK*}

\end{document}